\documentclass[11pt,a4paper]{article}

\usepackage[a4paper,margin=1in]{geometry}
\usepackage{amsmath,amssymb,amsthm}
\usepackage{bm}
\usepackage{physics}
\usepackage{slashed}
\usepackage{graphicx}
\usepackage{float}
\usepackage{booktabs}
\usepackage{multirow}
\usepackage{array}
\usepackage{hyperref}
\usepackage{xcolor}
\usepackage{cite}
\usepackage{caption}
\usepackage{subcaption}
\usepackage{authblk}

\hypersetup{
    colorlinks=true,
    linkcolor=blue,
    citecolor=blue,
    urlcolor=blue
}

\newcommand{\mnu}{M_\nu}
\newcommand{\brmue}{\mathrm{BR}(\mu \to e\gamma)}

\newcommand{\ohm}{\Omega h^2}

\title{
\textbf{
Emergent Neutrino Texture Geometry from Dark Matter and Lepton Flavor Violation in the Scotogenic Model
}
}

\author{
Avinanda Chaudhuri\\
\small Department of Physics,\\ Brahmananda Keshab Chandra College,\\ 111/2, B. T. Road, India - 700108\\
\small \thanks{\texttt{aviphys@gmail.com}}
}

\date{}

\begin{document}

\maketitle

\begin{abstract}

We investigate the emergence of approximate neutrino texture structures in the minimal scotogenic model through large-scale Casas--Ibarra parameter scans subject to lepton flavor violation and dark matter constraints. We demonstrate that approximate suppressions can dynamically emerge from phenomenological consistency conditions. The interplay between relic density requirements, radiative neutrino mass generation, and lepton flavor violating observables induces a nontrivial flavor geometry in parameter space. Particular suppressions in the $(e\mu)$ and $(e\tau)$ sectors arise naturally, while diagonal entries strongly resist cancellation. We further compare normal and inverted mass hierarchies, analyze reduced versus full Casas--Ibarra geometries, and identify approximate scaling relations linking dark matter and flavor observables. Our results suggest that emergent flavor structures may represent dynamical consequences of radiative neutrino mass generation rather than externally imposed flavor symmetries.

\end{abstract}

\section{Introduction}

The discovery of neutrino oscillations established that neutrinos are massive and mixed particles, thereby providing direct evidence for physics beyond the Standard Model. Over the past decades, oscillation experiments~\cite{T2K, NOvA, SuperK, DayaBay, RENO, DoubleChooz, SNO, KamLAND, DUNE, HyperK, JUNO, NuFIT} have measured the neutrino mass-squared differences and leptonic mixing angles with remarkable precision. In contrast to the quark sector, the leptonic mixing matrix exhibits large mixing angles, suggesting that the underlying flavor structure of neutrinos may originate from fundamentally different dynamics. Despite substantial experimental progress, the origin of neutrino masses and the structure of the leptonic flavor sector remain unresolved problems in particle physics.

One of the most widely explored approaches to understanding neutrino flavor structure involves the study of neutrino mass matrix textures. Texture-zero frameworks~\cite{FGM, XingTexture, Kageyama, FritzschXing, SinghTexture, DevTexture, ModularA4, LiaoMarfatia, Lashin} assume that specific entries of the neutrino mass matrix vanish exactly due to underlying flavor symmetries, leading to predictive relations among oscillation parameters, CP phases, and neutrino mass hierarchies. Such analyses have been extensively studied in both Majorana and Dirac neutrino frameworks~\cite{LudlDirac, BrancoTexture, FourZeroDirac, Adhikary, RaidalStrumia, Meloni}. However, most texture studies impose flavor structures \emph{a priori}, typically through discrete or continuous flavor symmetries. In contrast, comparatively little attention has been devoted to the possibility that approximate texture structures may emerge dynamically from phenomenological consistency conditions themselves.

Radiative neutrino mass generation models~\cite{Zee, Babu, CaiReview, Bonnet, Sierra, ScotogenicReview, KNT, IDM, Hambye} provide a particularly attractive setting for exploring such emergent flavor structures. Among these, the scotogenic model offers a minimal and elegant framework in which neutrino masses arise radiatively at one-loop order while simultaneously incorporating a viable dark matter candidate~\cite{MaScotogenic}. The model extends the Standard Model particle content through an inert scalar doublet and singlet fermions charged under a discrete $Z_2$ symmetry. In this framework, the same Yukawa couplings responsible for generating neutrino masses also induce charged lepton flavor violating processes such as $\mu \to e\gamma$ and contribute directly to dark matter annihilation dynamics~\cite{Kubo, LopezHonorez, TomaVicente, Restrepo, MolinaroFIMP, HirschFIMP, VicenteReview}. Consequently, neutrino flavor structure, lepton flavor violation, and dark matter phenomenology become intrinsically interconnected.

This interplay motivates a natural question:
\begin{center}
\emph{
Can viable scotogenic parameter space dynamically generate approximate neutrino texture structures without imposing flavor symmetries by hand?
}
\end{center}

This question is nontrivial because several competing effects simultaneously shape the flavor geometry of the model. Lepton flavor violation constraints strongly suppress flavor-violating Yukawa structures, while dark matter relic density considerations typically favor sizable interactions. In addition, the Casas--Ibarra parametrization introduces geometric cancellations through complex orthogonal rotations, and the resulting interference structure depends sensitively on the neutrino mass ordering. 

In this work, we investigate the emergence of approximate neutrino texture suppressions in the minimal scotogenic model through large-scale Casas--Ibarra parameter scans subject to lepton flavor violation and dark matter constraints. We do not impose vanishing matrix elements externally. Instead, we quantify dynamically generated suppressions through emergent texture measures and analyze their statistical distributions across parameter space. We compare reduced and full Casas--Ibarra geometries, study both normal and inverted neutrino mass hierarchies, investigate dark matter--LFV tension structures, and derive approximate scaling relations linking flavor and dark matter observables.

The paper is organized as follows. In Sec.~II, we review the minimal scotogenic framework and the relevant neutrino mass, dark matter, and LFV observables. In Sec.~III, we introduce the Casas--Ibarra parametrization, define the emergent texture suppression measures used throughout the analysis and our numerical strategy. Section~IV presents the emergent texture structures obtained from numerical scans, while Sec.~V compares normal and inverted neutrino mass hierarchies. We discuss the analytic origin of emergent neutrino textures in Sec.~VI. In Sec.~VII, we investigate the dark matter--LFV geometric tension and analyze viable regions of parameter space. Section~VIII describes emergent hierarchy invariant and flavor compression. Finally, Sec.~IX summarizes our conclusions and discusses future directions.

\section{Minimal Scotogenic Framework}

The framework extends the Standard Model through the addition of three singlet fermions $N_i$ $(i=1,2,3)$ and an inert scalar doublet
\begin{equation}
\eta =
\begin{pmatrix}
\eta^+ \\
\eta^0
\end{pmatrix},
\end{equation}
all of which are odd under an imposed discrete $Z_2$ symmetry, whereas the Standard Model fields remain $Z_2$ even. As a consequence of the exact $Z_2$ symmetry, tree-level neutrino masses are forbidden and the lightest $Z_2$-odd particle becomes stable, thereby providing a dark matter candidate.

The relevant Yukawa interactions are given by
\begin{equation}
\mathcal{L}_Y
=
Y_{\alpha i}\,
\overline{L_\alpha}\,
\widetilde{\eta}\,
N_i
+
\frac12 M_i
\overline{N_i^c}N_i
+
\mathrm{h.c.},
\end{equation}
where $L_\alpha$ $(\alpha=e,\mu,\tau)$ denote the Standard Model lepton doublets, $Y_{\alpha i}$ are the neutrino Yukawa couplings, and $M_i$ represent the singlet fermion Majorana masses. The inert scalar doublet does not acquire a vacuum expectation value, thereby preserving the discrete symmetry after electroweak symmetry breaking.

The scalar potential of the model can be written as
\begin{align}
V(H,\eta)
&=
\mu_1^2 |H|^2
+
\mu_2^2 |\eta|^2
+
\lambda_1 |H|^4
+
\lambda_2 |\eta|^4
\nonumber\\
&\quad
+
\lambda_3 |H|^2 |\eta|^2
+
\lambda_4 |H^\dagger \eta|^2
+
\frac{\lambda_5}{2}
\left[
(H^\dagger \eta)^2 + \mathrm{h.c.}
\right],
\end{align}
where the parameter $\lambda_5$ plays a crucial role in generating nonzero neutrino masses through loop effects.

After electroweak symmetry breaking, the neutral inert scalar component splits into its CP-even and CP-odd components,
\begin{equation}
\eta^0
=
\frac1{\sqrt2}
(\eta_R + i\eta_I),
\end{equation}
with masses
\begin{align}
m_R^2
&=
m_\eta^2
+
(\lambda_3+\lambda_4+\lambda_5)v^2,
\\
m_I^2
&=
m_\eta^2
+
(\lambda_3+\lambda_4-\lambda_5)v^2.
\end{align}

The scalar mass splitting directly controls the radiative neutrino mass scale and strongly affects the flavor geometry of the model.
\begin{equation}
\Delta m
\equiv
m_R-m_I
\end{equation}

\subsection{Radiative Neutrino Masses}

Light neutrino masses are generated radiatively at one-loop order through the exchange of inert scalars and singlet fermions. The resulting neutrino mass matrix is given by
\begin{equation}
(\mnu)_{\alpha\beta}
=
\sum_i
\frac{
Y_{\alpha i}Y_{\beta i}M_i
}{
32\pi^2
}
\left[
\frac{m_R^2}{m_R^2-M_i^2}
\ln\left(
\frac{m_R^2}{M_i^2}
\right)
-
\frac{m_I^2}{m_I^2-M_i^2}
\ln\left(
\frac{m_I^2}{M_i^2}
\right)
\right].
\label{eq:scotogenic_mass}
\end{equation}

It is convenient to rewrite Eq.~(\ref{eq:scotogenic_mass}) in compact matrix notation as
\begin{equation}
\mnu
=
Y^T \Lambda Y,
\label{eq:master_mass}
\end{equation}
where $\Lambda$ denotes the diagonal loop matrix,

\begin{equation}
\Lambda_i
=
\frac{
M_i
}{
32\pi^2
}
\left[
\frac{m_R^2}{m_R^2-M_i^2}
\ln\left(
\frac{m_R^2}{M_i^2}
\right)
-
\frac{m_I^2}{m_I^2-M_i^2}
\ln\left(
\frac{m_I^2}{M_i^2}
\right)
\right].
\end{equation}

In the small splitting limit,
\begin{equation}
|m_R^2-m_I^2|
\ll
m_\eta^2,
\end{equation}
the neutrino mass matrix approximately scales as
\begin{equation}
\mnu
\propto
Y^2 \Delta m,
\label{eq:scaling_neutrino}
\end{equation}
revealing an important connection between Yukawa magnitudes and scalar splitting.

The same Yukawa couplings responsible for radiative neutrino mass generation also induce charged lepton flavor violating processes. Among these, the radiative decay
\begin{equation}
\mu \to e\gamma
\end{equation}
provides one of the strongest phenomenological constraints~\cite{MEG, MEGII} on the model parameter space.

The corresponding branching ratio~\cite{MEGA, Hisano, KunoOkada, CalibbiSignorelli} is approximately given by
\begin{equation}
\brmue
=
\frac{
3\alpha_{\mathrm{em}}
}{
64\pi G_F^2 m_\eta^4
}
\left|
\sum_i
Y_{ei}Y^*_{\mu i}
F\left(
\frac{M_i^2}{m_\eta^2}
\right)
\right|^2,
\label{eq:lfv}
\end{equation}
where $F(x)$ denotes the standard loop function,
\begin{equation}
F(x)
=
\frac{
1-6x+3x^2+2x^3-6x^2\ln x
}{
6(1-x)^4
}.
\end{equation}

Equation~(\ref{eq:lfv}) demonstrates that LFV observables are directly sensitive to flavor off-diagonal combinations of the Yukawa matrix. Consequently, LFV constraints strongly influence the emergent flavor geometry of the neutrino mass matrix.

\subsection{Fermionic Dark Matter}

In the present work we focus on the fermionic dark matter realization of the scotogenic model~\cite{Suematsu,Klasen, Hugle, Dolle, Rojas, Borah, YagunaFreezeIn, LindnerScoto }, where the lightest singlet fermion $N_1$ constitutes the dark matter candidate. The relic abundance is determined primarily through annihilation processes of the form
\begin{equation}
N_1 N_1
\to
\ell_\alpha \ell_\beta,
\end{equation}
mediated by inert scalar exchange.

At the scaling level, the thermally averaged annihilation cross section behaves approximately as
\begin{equation}
\langle \sigma v \rangle
\sim
\frac{
|Y|^4 M_1^2
}{
(m_\eta^2+M_1^2)^2
},
\label{eq:sigmav}
\end{equation}
leading to the approximate relic density relation
\begin{equation}
\ohm
\propto
\frac1{\langle \sigma v \rangle}.
\label{eq:relic}
\end{equation}

Here, $\Omega h^2$ is the present-day dark matter relic density, where $\Omega = \rho_{\rm DM}/\rho_c$ denotes the dark matter density parameter and $h$ is the reduced Hubble constant.
The quantity $ \langle \sigma v \rangle $ corresponds to the thermally averaged dark matter annihilation cross section times relative velocity.

Equations~(\ref{eq:lfv}) and (\ref{eq:relic}) immediately reveal the origin of the dark matter--LFV tension. Increasing Yukawa couplings enhances dark matter annihilation and lowers the relic density, but simultaneously increases LFV amplitudes. Conversely, suppressing LFV generally drives the model toward dark matter overabundance.

\section{Casas--Ibarra Geometry and Emergent Texture Measures}

To explore the full flavor geometry compatible with oscillation data, we employ the Casas--Ibarra parametrization~\cite{CasasIbarra} adapted to radiative neutrino mass generation. The starting point is the radiative neutrino mass relation, Eq.(~\ref{eq:master_mass}). The light neutrino mass matrix can be diagonalized as
\begin{equation}
U^\dagger \mnu U^*
=
D_\nu,
\end{equation}
with
\begin{equation}
D_\nu
=
\mathrm{diag}(m_1,m_2,m_3),
\end{equation}
and $U$ denoting the PMNS mixing matrix~\cite{Pontecorvo, MNS, PDG}.

Following the Casas--Ibarra construction, the Yukawa matrix may be reconstructed as
\begin{equation}
Y
=
\sqrt{\Lambda^{-1}}\,
R\,
\sqrt{D_\nu}\,
U^\dagger,
\label{eq:casas}
\end{equation}
where $R$ is a complex orthogonal matrix satisfying
\begin{equation}
R^T R
=
\mathbb{I}.
\end{equation}

Equation~(\ref{eq:casas}) plays a crucial role throughout our analysis because it separates:
\begin{itemize}

\item experimentally measured neutrino parameters,

\item loop-induced mass scales,

\item and unconstrained flavor geometry encoded in $R$.

\end{itemize}

As a consequence, the Casas--Ibarra framework provides a natural arena for studying emergent flavor structures and cancellation mechanisms in the scotogenic model.

\subsection{Reduced and Full Orthogonal Geometry}

In the most general case, the complex orthogonal matrix $R$ may be parameterized using three independent complex angles,
\begin{equation}
R
=
R_{23}(z_{23})
R_{13}(z_{13})
R_{12}(z_{12}),
\end{equation}
where
\begin{equation}
z_{ij}
=
x_{ij}
+
i y_{ij}
\end{equation}
contain both real rotational components and imaginary hyperbolic deformations.

Explicitly, the elementary rotation matrices take the form
\begin{equation}
R_{12}(z_{12})
=
\begin{pmatrix}
\cos z_{12} & \sin z_{12} & 0 \\
-\sin z_{12} & \cos z_{12} & 0 \\
0 & 0 & 1
\end{pmatrix},
\end{equation}
with analogous expressions for $R_{13}$ and $R_{23}$.

The imaginary components $y_{ij}$ generate hyperbolic rotations,
\begin{equation}
\cos(x+iy)
=
\cos x \cosh y
-
i \sin x \sinh y,
\end{equation}
which can significantly amplify or suppress specific Yukawa structures. Consequently, the complex geometry of the Casas--Ibarra parametrization introduces highly nontrivial interference patterns in the reconstructed neutrino mass matrix.

In addition to the full orthogonal geometry, we also study a reduced parametrization involving a single complex angle,
\begin{equation}
R(z)
=
\begin{pmatrix}
0 & \cos z & \sin z \\
0 & -\sin z & \cos z \\
1 & 0 & 0
\end{pmatrix},
\label{eq:reduced_R}
\end{equation}
which is frequently employed in minimal radiative neutrino mass studies. One of the goals of this work is to determine whether the reduced geometry already captures the dominant emergent texture structures observed in the full parameter space.

\subsection{Emergent Texture Measures and Cancellation Geometry}
The neutrino mass matrix entries can be written explicitly as
\begin{equation}
(\mnu)_{\alpha\beta}
=
\sum_i
m_i
U_{\alpha i}
U_{\beta i},
\label{eq:decomposition}
\end{equation}
revealing that approximate texture structures originate from interference among different mass-eigenstate contributions.

To quantify the strength of cancellations, we define the cancellation ratio
\begin{equation}
\epsilon_{\alpha\beta}
=
\frac{
|(\mnu)_{\alpha\beta}|
}{
\sum_i
|m_i U_{\alpha i}U_{\beta i}|
}.
\label{eq:cancellation}
\end{equation}

Values
\begin{equation}
\epsilon_{\alpha\beta}\ll1
\end{equation}
indicate strong destructive interference, whereas
\begin{equation}
\epsilon_{\alpha\beta}\sim1
\end{equation}
correspond to coherent addition of contributions.

This decomposition proves particularly useful for understanding:
\begin{itemize}

\item why certain sectors naturally suppress,

\item why others strongly resist cancellation,

\item and how normal and inverted hierarchies generate qualitatively different flavor geometries.

\end{itemize}

\subsection{Numerical Strategy}

Our numerical analysis combines: oscillation-consistent neutrino reconstruction, complex Casas--Ibarra scans, LFV constraints, dark matter relic density requirements and texture suppression diagnostics.

\begin{table}[htbp]
\centering
\caption{Parameter ranges used in the Casas--Ibarra numerical scans.}
\label{tab:scan_ranges}
\begin{tabular}{ccc}
\hline
Parameter & Range & Description \\
\hline

$M_1$ & $300~\mathrm{GeV} - 3~\mathrm{TeV}$ & Fermion DM mass \\

$\Delta m$ & $1 - 400~\mathrm{GeV}$ & Scalar mass splitting \\

$x_{12,13,23}$ & $[-\pi,\pi]$ & Real Casas--Ibarra angles \\

$y_{12,13,23}$ & $[-5,5]$ & Imaginary Casas--Ibarra angles \\

$m_{\rm lightest}$ & $10^{-5} - 10^{-1}~\mathrm{eV}$ & Lightest neutrino mass \\

$\delta_{\rm CP}$ & $[0,2\pi]$ & Dirac CP phase \\

\hline
\end{tabular}
\end{table}

For each scan point, we reconstruct the Yukawa matrix through Eq.~(\ref{eq:casas}), compute the resulting radiative neutrino mass matrix, evaluate LFV observables and relic density constraints, and determine the emergent suppression measures defined in Eq.~(\ref{eq:cancellation}).

Both normal hierarchy (NH) and inverted hierarchy (IH) are investigated independently in order to identify hierarchy-dependent geometric structures. Throughout the analysis, we emphasize statistical distributions and emergent patterns rather than isolated benchmark points.

This approach allows us to study flavor textures as dynamically generated geometric structures arising from the interplay between:
\begin{equation*}
\text{oscillations}
\quad\oplus\quad
\text{LFV}
\quad\oplus\quad
\text{dark matter}
\quad\oplus\quad
\text{radiative reconstruction}.
\end{equation*}

\subsection{Numerical Scan Details}

The numerical analysis was performed using a modular Monte Carlo scan framework implemented in Python. Random parameter points were generated using both direct Yukawa sampling and Casas--Ibarra reconstruction techniques within the minimal fermionic scotogenic model.

For the Casas--Ibarra reconstruction scans, the heavy fermion masses were typically chosen in the TeV regime,
\[
M_i \sim \mathcal{O}(1~\mathrm{TeV}),
\]
with representative benchmark choices such as
\[
(M_1,M_2,M_3)
=
(1000,1500,2000)\,\mathrm{GeV},
\]
while extended scans also allowed fully randomized heavy spectra in the range
\[
300~\mathrm{GeV}
<
M_i
<
3000~\mathrm{GeV}.
\]
The inert scalar sector was scanned over
\[
500~\mathrm{GeV}
<
m_\eta
<
4000~\mathrm{GeV},
\]
with scalar mass splittings
\[
1~\mathrm{GeV}
<
\Delta m
<
400~\mathrm{GeV}.
\]

The lightest neutrino mass was sampled logarithmically according to
\[
10^{-5}\,\mathrm{eV}
<
m_{\rm lightest}
<
10^{-1}\,\mathrm{eV},
\]
while the complex Casas--Ibarra angles were parameterized as
\[
z_{ij}
=
x_{ij}
+
i y_{ij},
\]
with
\[
x_{ij}\in[-\pi,\pi],
\qquad
y_{ij}\in[-1.5,1.5].
\]

The Yukawa matrices were reconstructed numerically using the Casas--Ibarra parametrization and subsequently used to compute the radiative neutrino mass matrix, charged lepton flavor violating observables, and dark matter relic abundance. The branching ratio for the process
\[
\mu \rightarrow e\gamma
\]
was evaluated for each parameter point, and the current MEG bound
\[
\mathrm{BR}(\mu\to e\gamma)
<
4.2\times10^{-13}
\]
was imposed throughout the analysis. 

Texture suppressions were quantified through normalized matrix-entry measures of the form described in Eq.(~\ref{eq:cancellation}) allowing a systematic study of approximate texture-zero emergence across viable parameter space. 

Separate scans were performed for: direct Yukawa sampling, one-angle Casas--Ibarra reconstruction, full three-angle Casas--Ibarra geometry, dark matter plus LFV constrained scans, coannihilation-enhanced relic-density scans, flavor geometry manifold studies and hierarchy--entropy analyses.

Typical scan sizes ranged between
\[
\mathcal{O}(5\times10^3)
\quad \text{and} \quad
\mathcal{O}(5\times10^4)
\]
randomly generated parameter points depending on the analysis channel. 

For the dark matter analysis, %
only parameter points satisfying both LFV and relic-density constraints were retained. 

To characterize the emergent flavor structure statistically, geometric observables including the flavor entropy
\[
S_{\rm flavor}
=
-\sum_i p_i \ln p_i,
\]
and the hierarchy invariant
\[
\mathcal{H}
=
\frac{
\epsilon_{\mu\mu}
+
\epsilon_{\tau\tau}
}{
\epsilon_{e\mu}
+
\epsilon_{e\tau}
},
\]
were evaluated across the viable parameter manifold.

\section{Emergent Texture Structures}

\subsection{Statistical Texture Hierarchies}

\begin{figure}[H]
\centering
\includegraphics[width=0.95\textwidth]{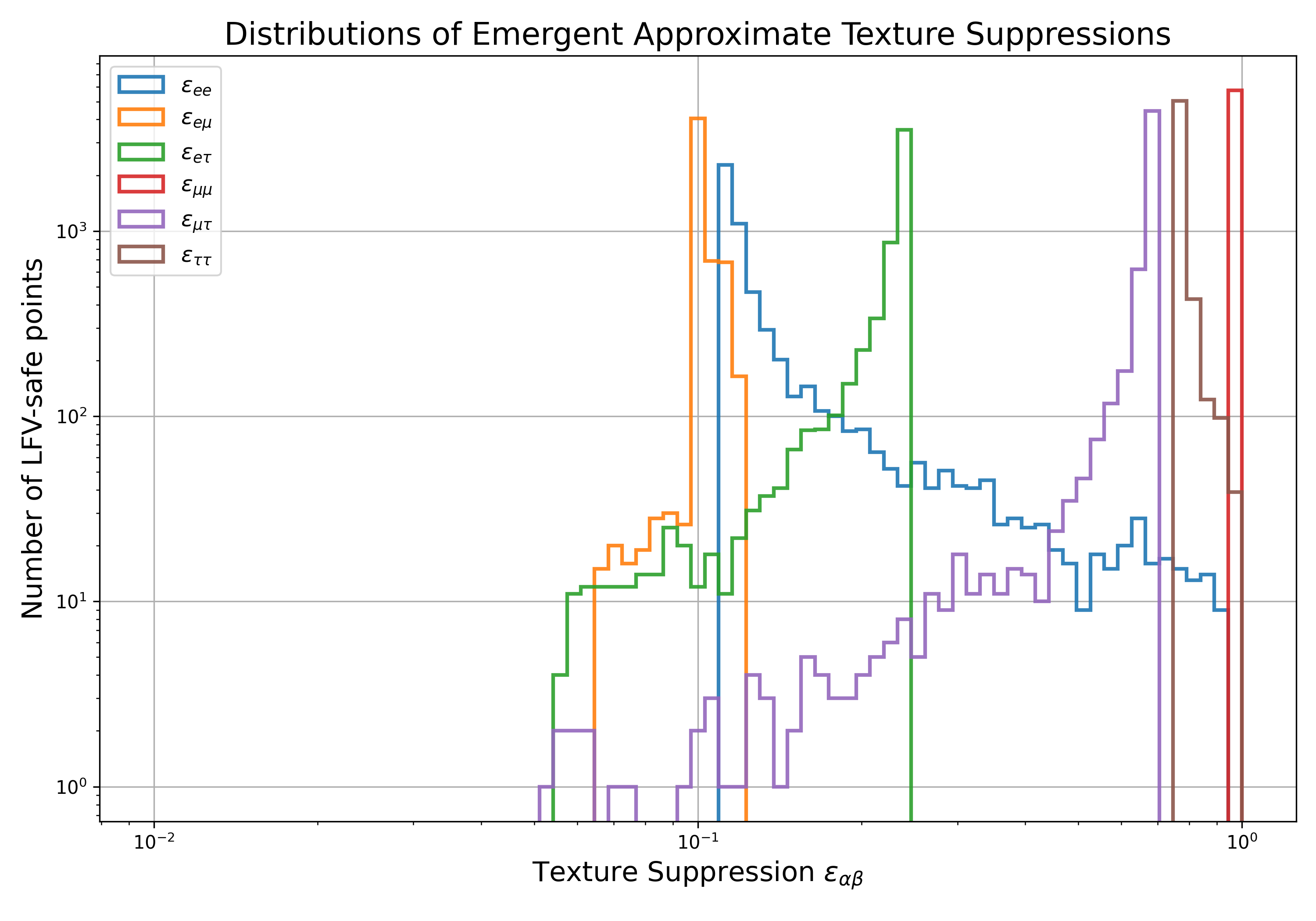}
\caption{
Statistical distributions of emergent texture suppression parameters $\epsilon_{\alpha\beta}$ obtained from LFV-safe parameter scans in the minimal scotogenic model. Off-diagonal electron sectors exhibit substantially stronger suppression than diagonal atmospheric sectors.
}
\label{fig:texture_histograms}
\end{figure}

Figure~\ref{fig:texture_histograms} shows the statistical distributions of the texture suppression parameters obtained from LFV-safe Casas--Ibarra scans. Several important features immediately emerge.

Most notably, the off-diagonal electron sectors
\begin{equation}
M_{e\mu},
\qquad
M_{e\tau}
\end{equation}
display significantly enhanced accessibility to approximate suppression. In contrast, the diagonal atmospheric sectors
\begin{equation}
M_{\mu\mu},
\qquad
M_{\tau\tau}
\end{equation}
strongly resist cancellation and remain concentrated near
\begin{equation}
\epsilon_{\alpha\beta}\sim1.
\end{equation}

This behavior indicates that the emergent flavor geometry of the scotogenic model is highly nonuniform. Approximate texture structures do not arise democratically across all matrix elements, but instead appear preferentially in specific flavor sectors.

\begin{table}[htbp]
\centering
\caption{Experimental constraints imposed throughout the numerical analysis.}
\label{tab:constraints}
\begin{tabular}{ccc}
\hline
Observable & Constraint & Reference \\
\hline

$\mathrm{BR}(\mu\to e\gamma)$ &
$< 4.2 \times 10^{-13}$ &
MEG \\

$\Omega h^2$ &
$0.12 \pm 0.01$ &
Planck \\

$\Delta m_{21}^2$ &
$7.4 \times 10^{-5}~\mathrm{eV}^2$ &
NuFIT \\

$|\Delta m_{31}^2|$ &
$2.5 \times 10^{-3}~\mathrm{eV}^2$ &
NuFIT \\

\hline
\end{tabular}
\end{table}

An especially important observation concerns the $(e\mu)$ sector, whose suppression distribution extends toward comparatively small values,
\begin{equation}
\epsilon_{e\mu}
\ll1,
\end{equation}
without any externally imposed flavor symmetry. This demonstrates that approximate flavor textures can emerge dynamically as a consequence of viable radiative reconstruction itself.

\subsection{Texture Suppression and LFV}

To understand the physical origin of these emergent structures, it is instructive to study the correlation between texture suppressions and lepton flavor violation observables.

\begin{figure}[H]
\centering
\includegraphics[width=0.8\textwidth]{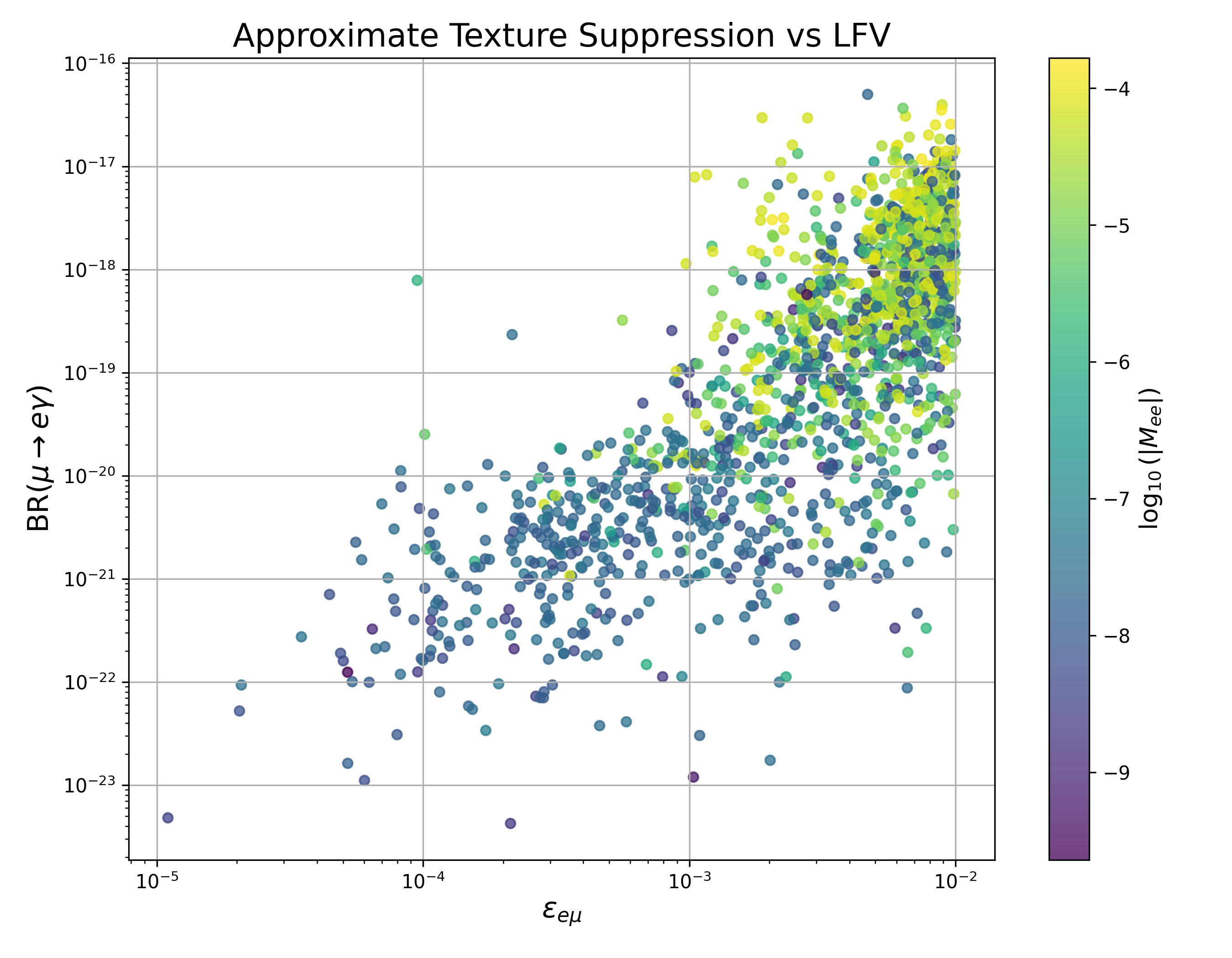}
\caption{
Correlation between the emergent suppression parameter $\epsilon_{e\mu}$ and the LFV observable $\mathrm{BR}(\mu\to e\gamma)$. Smaller values of $\epsilon_{e\mu}$ tend to populate LFV-safe regions more efficiently, indicating that approximate flavor suppressions emerge dynamically through LFV constraints.
}
\label{fig:texture_lfv}
\end{figure}

Figure~\ref{fig:texture_lfv} displays the relationship between the suppression parameter $\epsilon_{e\mu}$ and the branching ratio
\begin{equation}
\brmue.
\end{equation}

A clear geometric pattern becomes visible. LFV-safe regions preferentially populate smaller values of $\epsilon_{e\mu}$, indicating that approximate suppression in the $(e\mu)$ sector is strongly correlated with reduced flavor-violating amplitudes. Since the LFV observable depends directly on combinations of the form
\begin{equation}
\sum_i
Y_{ei}Y_{\mu i}^*,
\end{equation}
the suppression of the corresponding neutrino mass matrix entry naturally reduces LFV amplitudes through destructive interference.

This result is conceptually important because it demonstrates that approximate flavor textures are not arbitrary artifacts of parameter scanning.

\subsection{Reduced versus Full Geometry}

An important question concerns whether the emergent texture structures identified above depend strongly on the dimensionality of the Casas--Ibarra parametrization.

To investigate this issue, we compare scans performed using:
\begin{itemize}

\item the reduced one-angle orthogonal geometry,

\item and the full three-angle complex orthogonal parametrization.

\end{itemize}

\begin{figure}[H]
\centering
\includegraphics[width=0.9\textwidth]{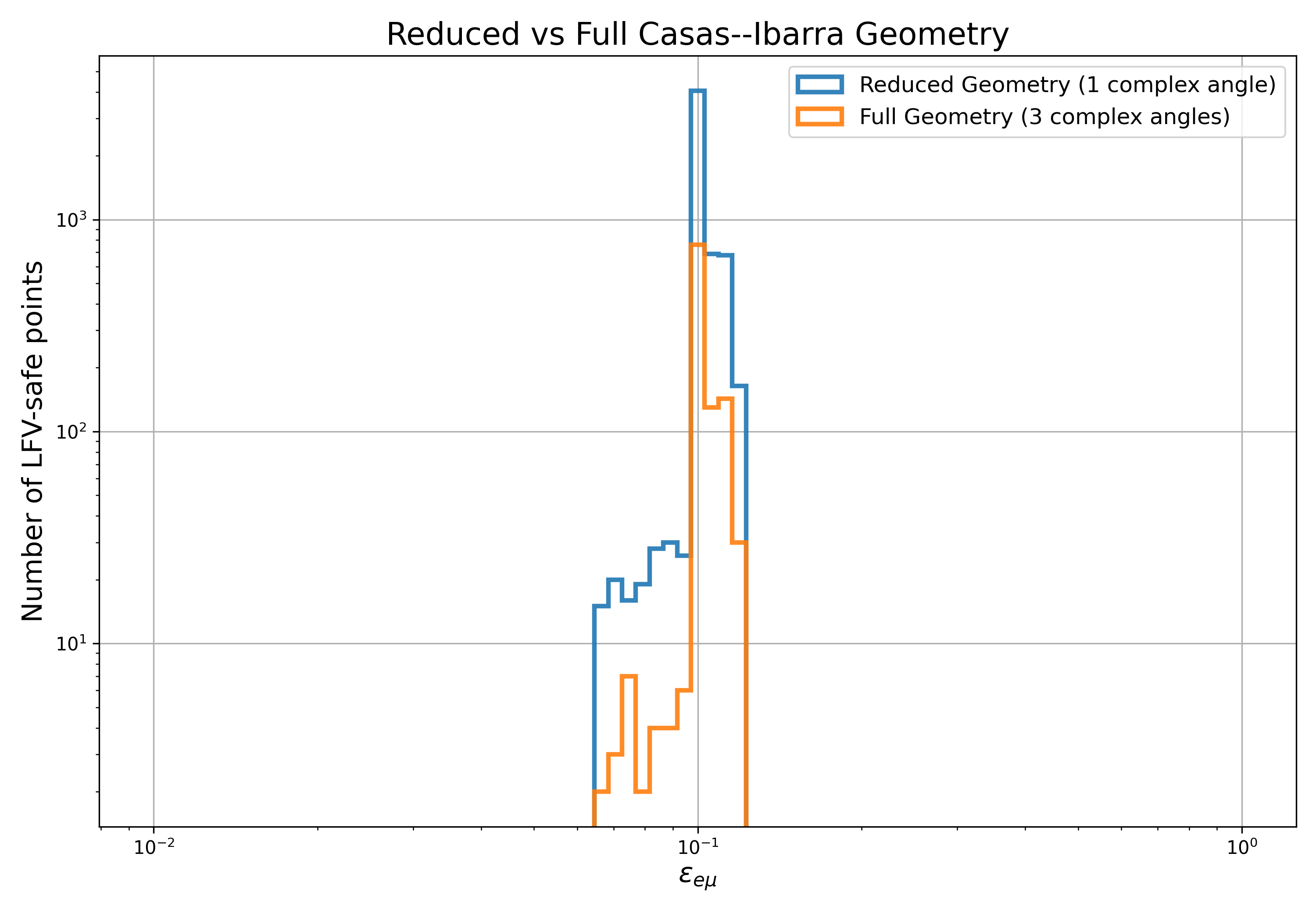}
\caption{
Comparison between reduced and full Casas--Ibarra geometries. The dominant emergent suppression structures remain qualitatively stable, indicating that the observed flavor hierarchies are robust geometric features rather than artifacts of a particular parametrization.
}
\label{fig:reduced_full}
\end{figure}

Figure~\ref{fig:reduced_full} compares the corresponding suppression distributions.

Remarkably, the dominant suppression structures remain qualitatively stable across both parametrizations. In particular, the preferential accessibility of the $(e\mu)$ and $(e\tau)$ sectors persists even within the reduced geometry. This suggests that the emergent flavor hierarchies identified in our analysis are robust geometric properties of the scotogenic reconstruction framework itself rather than accidental artifacts of higher-dimensional parameter freedom.

\section{Normal and Inverted Hierarchy Geometry}

The flavor structure of the neutrino mass matrix depends sensitively on the underlying neutrino mass hierarchy. Since the individual matrix elements arise from interference among different mass-eigenstate contributions, the relative hierarchy among neutrino masses can strongly modify the accessibility of approximate texture suppressions.

In this section, we compare the emergent texture geometry obtained for:
\begin{itemize}

\item normal hierarchy (NH),

\item inverted hierarchy (IH).

\end{itemize}

Our analysis reveals substantial qualitative differences between the two cases, indicating that the accessible flavor structures of the scotogenic model are strongly hierarchy dependent.

\subsection{Mass Ordering Setup}

For normal hierarchy, the light neutrino spectrum satisfies
\begin{equation}
m_1 < m_2 < m_3,
\end{equation}
with
\begin{align}
m_2
&=
\sqrt{
m_1^2+\Delta m_{21}^2
},
\\
m_3
&=
\sqrt{
m_1^2+\Delta m_{31}^2
}.
\end{align}

In contrast, inverted hierarchy is characterized by
\begin{equation}
m_3 < m_1 < m_2,
\end{equation}
with
\begin{align}
m_1
&=
\sqrt{
m_3^2+|\Delta m_{31}^2|
},
\\
m_2
&=
\sqrt{
m_3^2+\Delta m_{21}^2+|\Delta m_{31}^2|
}.
\end{align}

The different hierarchy structures modify the interference properties of Eq.~(\ref{eq:decomposition}) and therefore alter the emergent cancellation geometry of the neutrino mass matrix.

\subsection{Texture Accessibility in NH and IH}

\begin{figure}[H]
\centering
\includegraphics[width=0.95\textwidth]{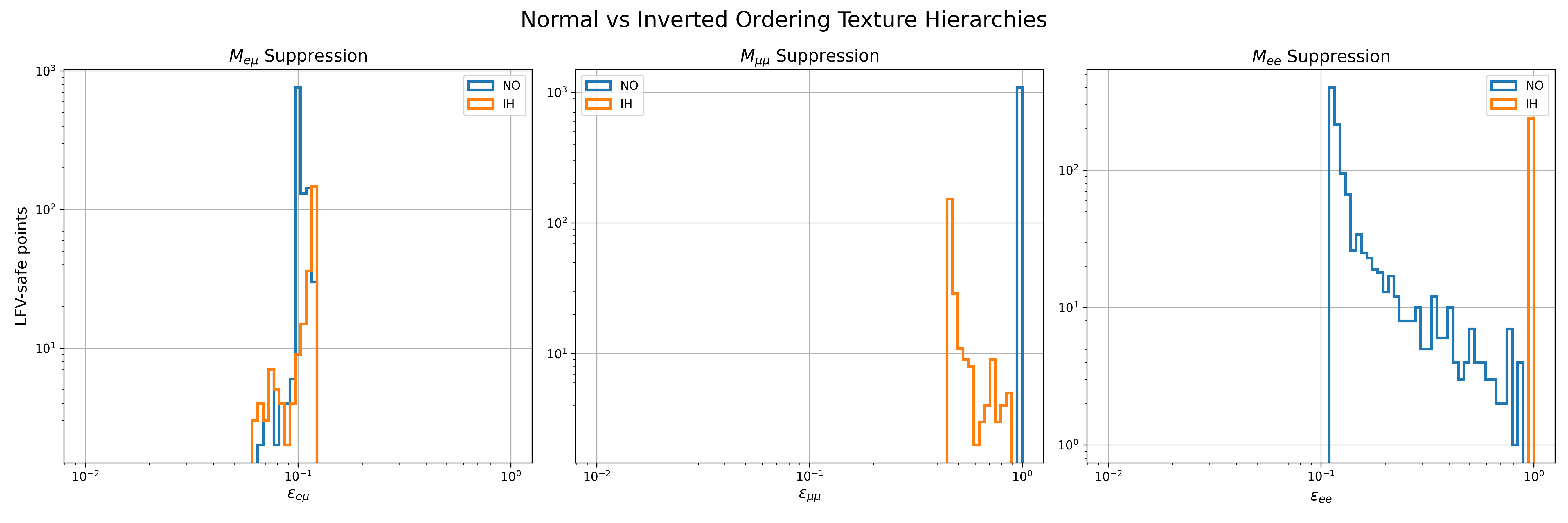}
\caption{
Comparison between emergent texture suppression distributions for normal hierarchy (NH) and inverted ordering (IH). Normal hierarchy permits significantly richer approximate texture structures, whereas inverted ordering strongly suppresses the accessibility of several cancellations, particularly in the $M_{ee}$ sector.
}
\label{fig:NO_IH}
\end{figure}

Figure~\ref{fig:NO_IH} compares the distributions of the suppression parameters $\epsilon_{\alpha\beta}$ obtained for normal and inverted neutrino mass hierarchy.

Several important differences immediately become apparent.

Most notably, normal hierarchy permits substantially stronger approximate suppressions across multiple flavor sectors. In particular, the distributions of
\begin{equation*}
\epsilon_{e\mu},
\qquad
\epsilon_{e\tau}
\end{equation*}
extend toward significantly smaller values in the NH case than in IH.

By contrast, inverted ordering strongly suppresses the accessibility of several approximate textures. This effect is especially pronounced in the $M_{ee}$ sector, where the distribution remains concentrated near $\epsilon_{ee}\sim1.$

This behavior indicates that the cancellation geometry associated with inverted ordering is substantially more constrained than in the normal hierarchy.

\section{Analytic Origin of Emergent Textures}


The numerical scans presented in the previous sections reveal a highly nonuniform accessibility
of approximate texture suppressions across flavor space. In particular, the off-diagonal electron
sectors
\begin{equation*}
M_{e\mu}, \qquad M_{e\tau}
\end{equation*}
frequently develop strong suppressions, whereas atmospheric diagonal sectors such as
\begin{equation*}
M_{\mu\mu}, \qquad M_{\tau\tau}
\end{equation*}
remain comparatively resistant to cancellation.

In this section, we derive the analytic origin of this hierarchy directly from the PMNS
interference structure of the neutrino mass matrix.

\subsection{Off-Diagonal Electron Sector}

We first consider the $(e\mu)$ element. Neglecting Majorana phases for simplicity and retaining
leading terms in $s_{13}$, one obtains
\begin{align}
M_{e\mu}
&=
m_1 U_{e1}U_{\mu1}
+
m_2 U_{e2}U_{\mu2}
+
m_3 U_{e3}U_{\mu3}
\nonumber\\[2mm]
&\simeq
-m_1 c_{12}s_{12}c_{23}
+
m_2 s_{12}c_{12}c_{23}
+
m_3 s_{13}s_{23}e^{-i\delta},
\label{eq:Me_mu_expansion}
\end{align}
where we have neglected subleading corrections proportional to $s_{13}m_{1,2}$.

For normal hierarchy,
\begin{equation*}
m_3 \gg m_2 > m_1,
\end{equation*}

the dominant contributions arise from the second and third terms in Eq.~(\ref{eq:Me_mu_expansion}).
Approximate suppression therefore requires
\begin{equation}
m_2 s_{12}c_{12}c_{23}
\sim
m_3 s_{13}s_{23},
\label{eq:Me_mu_cancellation_condition}
\end{equation}
together with an appropriate relative phase alignment controlled by the Dirac CP phase $\delta$.

Equation~(\ref{eq:Me_mu_cancellation_condition}) is particularly important because the observed
oscillation parameters naturally place the two contributions within the same order of magnitude:
\begin{align}
m_2 s_{12}c_{12}c_{23}
&\sim
{\cal O}(10^{-3}\text{--}10^{-2}) \text{ eV},
\\[2mm]
m_3 s_{13}s_{23}
&\sim
{\cal O}(10^{-3}\text{--}10^{-2}) \text{ eV}.
\end{align}

As a consequence, destructive interference becomes statistically accessible across broad regions
of parameter space without requiring extreme fine tuning.

This explains why the numerical scans naturally produce
\begin{equation*}
\epsilon_{e\mu} \ll 1
\end{equation*}
within large regions of LFV-safe parameter space.

An analogous argument applies to the $(e\tau)$ sector,
\begin{equation}
M_{e\tau}
\simeq
m_2 s_{12}c_{12}s_{23}
-
m_3 s_{13}c_{23}e^{-i\delta},
\end{equation}
which exhibits a similarly favorable interference structure.

Consequently, the off-diagonal electron sectors possess intrinsically enhanced cancellation
freedom due to the interplay between solar and atmospheric contributions.

\subsection{Atmospheric Diagonal Sector}

The situation differs substantially for the atmospheric diagonal elements. Consider the
$(\mu\mu)$ entry:
\begin{align}
M_{\mu\mu}
&=
m_1 U_{\mu1}^2
+
m_2 U_{\mu2}^2
+
m_3 U_{\mu3}^2
\nonumber\\[2mm]
&\simeq
m_1 s_{12}^2 c_{23}^2
+
m_2 c_{12}^2 c_{23}^2
+
m_3 s_{23}^2.
\label{eq:Mmu_mu_expansion}
\end{align}

Unlike the off-diagonal electron sector, all dominant contributions in
Eq.~(\ref{eq:Mmu_mu_expansion}) enter with the same sign at leading order.
As a consequence, the atmospheric diagonal sector is dominated by coherent addition rather than
destructive interference.

For normal hierarchy,
\begin{equation}
m_3 s_{23}^2
\end{equation}
typically provides the largest contribution, while the lighter-state terms remain positive and
therefore reinforce the overall matrix element.

Strong suppression would therefore require highly nontrivial phase correlations among multiple
subleading contributions, rendering cancellation statistically disfavored.

This explains why the numerical scans consistently produce
\begin{equation*}
\epsilon_{\mu\mu}
\sim
{\cal O}(1),
\end{equation*}
indicating strong resistance against approximate texture formation in the atmospheric diagonal
sector.

A similar argument applies to the $(\tau\tau)$ element.

\subsection{Normal versus Inverted Hierarchy}

The hierarchy dependence identified numerically can also be understood analytically.

For normal hierarchy,
\begin{equation*}
m_3 \gg m_2 > m_1,
\end{equation*}
the strong hierarchy among mass eigenstates permits relatively flexible interference patterns.
The lighter eigenstates can partially compensate the dominant atmospheric contribution,
allowing substantial cancellation freedom in the off-diagonal sectors.

In contrast, inverted ordering is characterized by
\begin{equation*}
m_1 \simeq m_2 \gg m_3,
\end{equation*}
leading to significantly more coherent interference structure.

For example, the $(ee)$ element becomes
\begin{equation}
M_{ee}
\simeq
m_1 c_{12}^2
+
m_2 s_{12}^2 e^{i\alpha},
\label{eq:Mee_IH}
\end{equation}
where $\alpha$ denotes the relevant Majorana phase.

Since the two dominant contributions possess comparable magnitudes,
strong suppression requires highly tuned relative phases,
\begin{equation}
\alpha \simeq \pi,
\end{equation}
together with near-maximal cancellation between the solar components.

Such configurations occupy only a comparatively small region of parameter space, explaining
why approximate $M_{ee}$ suppression becomes statistically inaccessible within the inverted
hierarchy.

The numerical results obtained in the previous sections confirm precisely this analytic
expectation.

\subsection{Emergent Geometric Interpretation}

The preceding derivations demonstrate that the observed texture hierarchies originate directly
from the interference geometry of the PMNS reconstruction itself.

\begin{table}[htbp]
\centering
\caption{Summary of emergent geometric behavior observed across viable parameter space.}
\label{tab:geometry_summary}
\begin{tabular}{ccc}
\hline
Observable & Normal Ordering & Inverted Ordering \\
\hline

$\epsilon_{e\mu}$ accessibility &
Strong suppression &
Moderate suppression \\

$\epsilon_{ee}$ suppression &
Accessible &
Strongly constrained \\

$\mathcal H$ &
Large &
Moderate \\

$S_{\rm flavor}$ &
Broader manifold &
Compressed manifold \\

DM--LFV tension &
Severe &
Severe \\

\hline
\end{tabular}
\end{table}

In particular:
\begin{itemize}
\item off-diagonal electron sectors possess naturally balanced competing contributions,

\item atmospheric diagonal sectors are dominated by coherent addition,

\item and inverted ordering substantially restricts cancellation freedom through quasi-degenerate eigenstate structure.
\end{itemize}

Consequently, the emergent flavor hierarchies identified throughout this work are not accidental
artifacts of numerical scanning, but instead represent intrinsic geometric properties of radiative
neutrino mass reconstruction within the scotogenic framework.
%
\subsection{Implications for Flavor Geometry}

The hierarchy dependence identified above has important conceptual implications.

First, it demonstrates that viable flavor geometry in the scotogenic model is strongly shaped by oscillation data itself. The experimentally preferred mass ordering directly influences which approximate textures can emerge dynamically.

Second, the qualitative asymmetry between NH and IH suggests that texture accessibility is fundamentally a geometric property of interference structure rather than merely a consequence of parameter counting.

Third, the stronger suppression accessibility observed in normal ordering indicates that NH naturally accommodates richer flavor hierarchies and more flexible cancellation patterns within the Casas--Ibarra reconstruction framework.

Taken together, these observations support the broader interpretation advocated throughout this work:
\begin{center}
\emph{
approximate neutrino textures can arise dynamically from the geometry of viable parameter space rather than externally imposed flavor symmetries.
}
\end{center}

In the following section, we extend this analysis by incorporating dark matter relic density constraints and investigating the resulting geometric tension between dark matter viability and lepton flavor violation.

\section{Dark Matter--LFV Geometric Tension}

In this section, we investigate the resulting geometric tension and analyze how it influences the emergent texture structures identified previously.

Our analysis reveals that the dark matter and LFV sectors impose mutually competing requirements across large regions of parameter space, leading to a strong dynamical obstruction within the minimal fermionic scotogenic model.

\subsection{Analytic Scaling Structure and the Origin of Tension}

The essential origin of the dark matter--LFV tension can already be understood at the scaling level.

The LFV branching ratio approximately behaves as
\begin{equation}
\brmue
\sim
\frac{
|Y|^4
}{
m_\eta^4
},
\label{eq:lfv_scaling}
\end{equation}
whereas the dark matter annihilation cross section scales as
\begin{equation}
\langle \sigma v \rangle
\sim
\frac{
|Y|^4 M_1^2
}{
(m_\eta^2+M_1^2)^2
}.
\label{eq:ann_scaling}
\end{equation}

Since the relic abundance is inversely proportional to the annihilation cross section,
\begin{equation}
\ohm
\propto
\frac1{\langle \sigma v \rangle},
\end{equation}
one obtains schematically
\begin{equation}
\ohm
\sim
\frac{
(m_\eta^2+M_1^2)^2
}{
|Y|^4 M_1^2
}.
\label{eq:omega_scaling}
\end{equation}

Equations~(\ref{eq:lfv_scaling}) and (\ref{eq:omega_scaling}) immediately reveal the competing tendencies:
\begin{itemize}

\item increasing Yukawa couplings lowers the relic density,

\item but simultaneously enhances LFV amplitudes.

\end{itemize}

Conversely:
\begin{itemize}

\item suppressing LFV requires smaller effective flavor couplings,

\item which generically drives the model toward dark matter overabundance.

\end{itemize}

This competition generates a highly constrained viable parameter manifold.

\subsection{Dark Matter versus LFV}

\begin{figure}[H]
\centering
\includegraphics[width=0.8\textwidth]{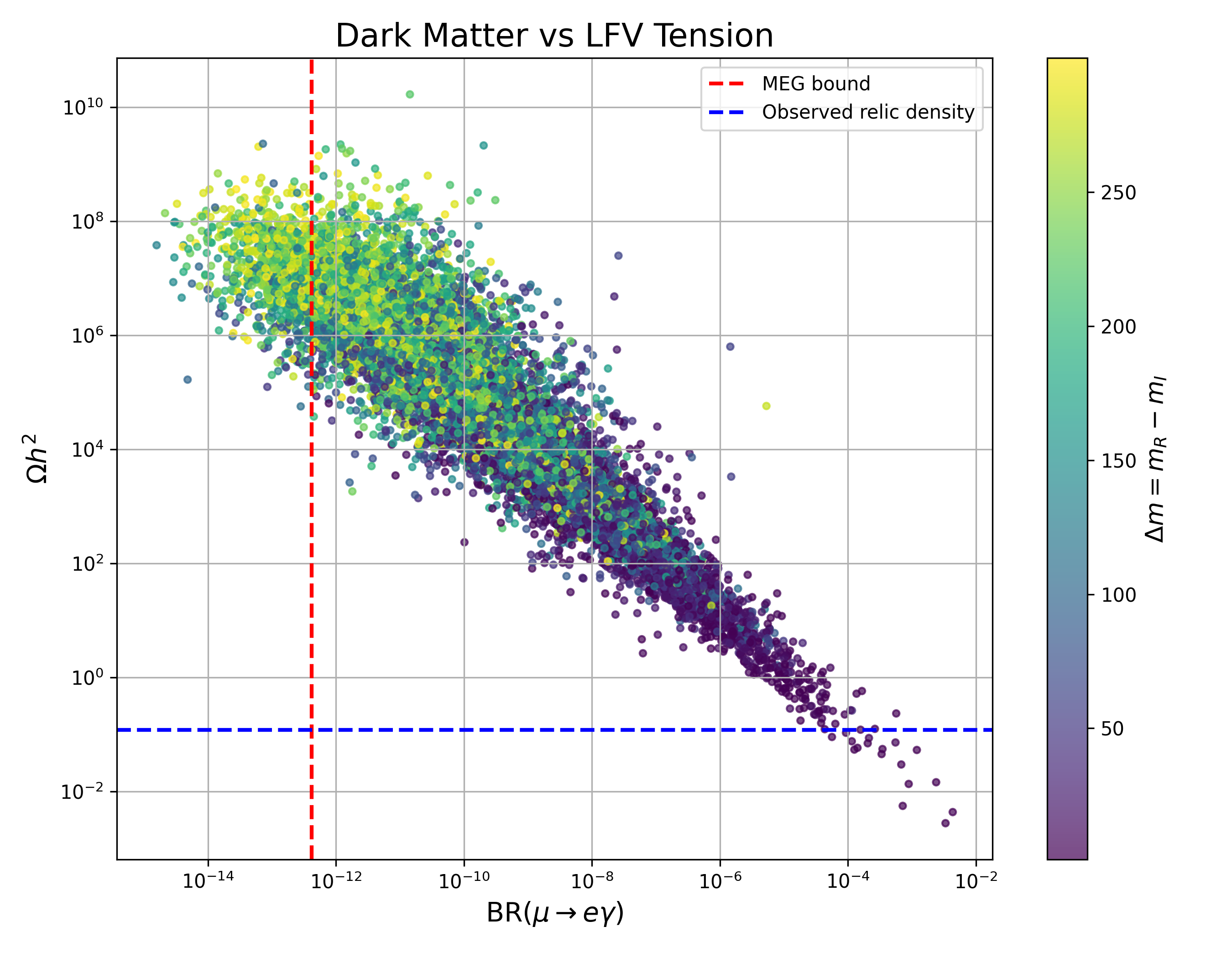}
\caption{
Dark matter relic abundance versus $\mathrm{BR}(\mu\to e\gamma)$ in the minimal fermionic scotogenic model. The color coding represents the scalar mass splitting $\Delta m = m_R-m_I$. LFV-safe regions generically tend toward dark matter overabundance, revealing a strong geometric tension between the two sectors.
}
\label{fig:dm_lfv_tension}
\end{figure}

The global structure of this tension is illustrated in Fig.~\ref{fig:dm_lfv_tension}, where we display the relic abundance as a function of the LFV branching ratio.

Several important features emerge from the figure.

First, the relic density and LFV observables exhibit a pronounced anti-correlation:
\begin{equation}
\ohm \downarrow
\qquad
\Longleftrightarrow
\qquad
\brmue \uparrow.
\end{equation}

This behavior directly reflects the common Yukawa dependence discussed above.

Second, LFV-safe regions preferentially populate the large relic-density regime,
\begin{equation}
\ohm \gg 0.12,
\end{equation}
indicating a strong tendency toward dark matter overclosure once flavor constraints are imposed.

Third, the scalar splitting
\begin{equation}
\Delta m
=
m_R-m_I
\end{equation}
plays a central role in shaping the tension structure. Smaller splittings generally require larger Yukawa couplings in order to reproduce the observed neutrino masses, thereby enhancing annihilation efficiency but simultaneously increasing LFV amplitudes. Larger splittings instead permit smaller Yukawas, suppressing LFV while worsening relic overabundance.

This competition demonstrates that the scalar sector acts as a dynamical control parameter for the flavor geometry of the model.

\subsection{Benchmark Dark Matter Validation with micrOMEGAs}

To further validate the qualitative dark matter behavior inferred from the scaling-level and geometric analyses, representative benchmark configurations were analyzed using \texttt{micrOMEGAs~6.3.0}~\cite{micromegas6}. The numerical evaluation includes thermal relic-density computation, dominant annihilation channels, and direct-detection constraints for selected effective benchmark points.

\begin{figure}[t]
\centering
\includegraphics[width=0.75\textwidth]{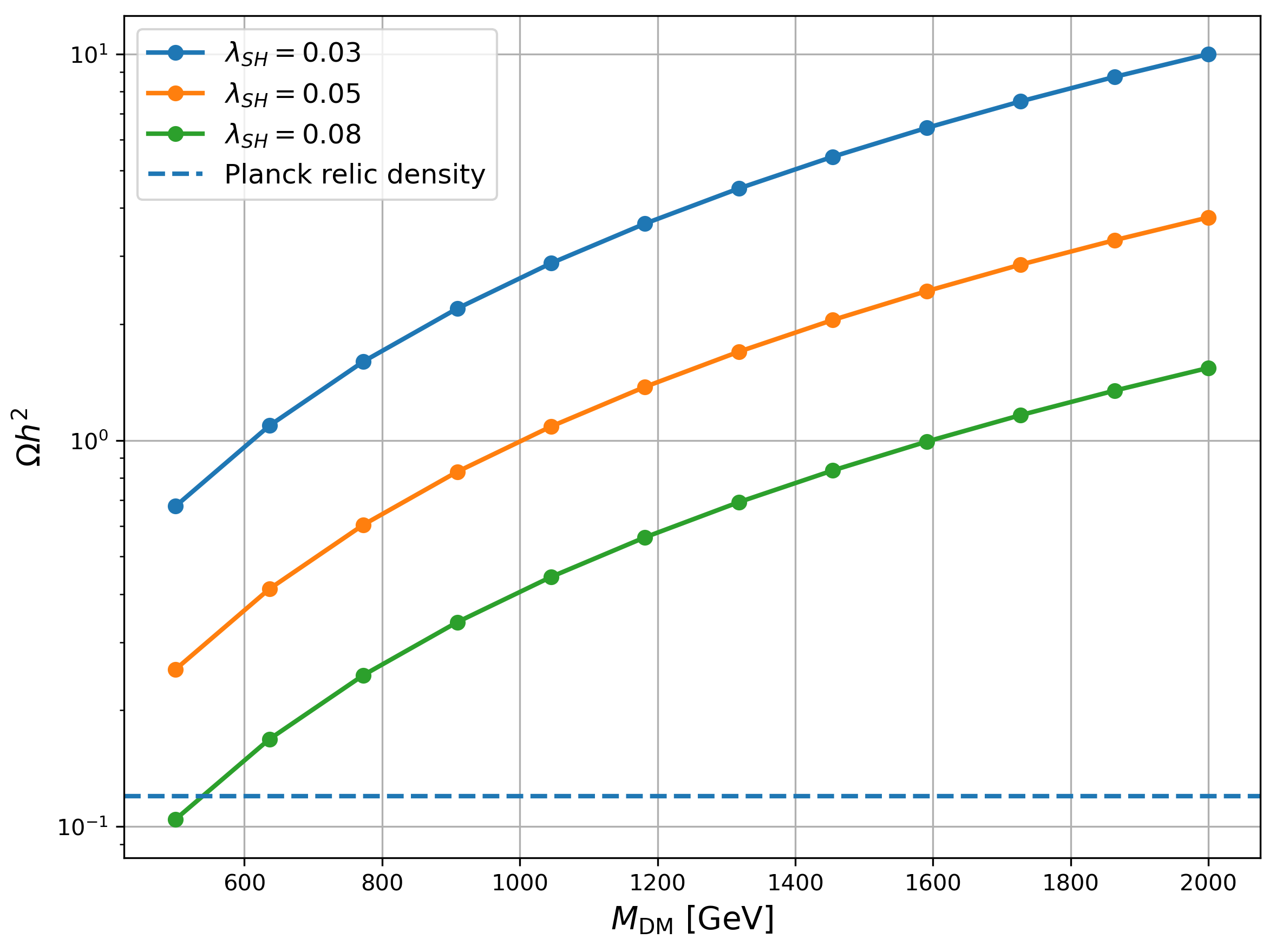}
\caption{
Dark matter relic density as a function of the dark matter mass for representative values of the effective Higgs-portal coupling $\lambda_{SH}$, obtained using \texttt{micrOMEGAs~6.3.0}. Larger portal couplings enhance annihilation efficiency and consequently reduce the relic abundance through more efficient thermal annihilation into electroweak final states. The horizontal dashed line corresponds to the observed Planck relic-density value, $\Omega h^2 \simeq 0.12$. The figure illustrates the emergence of overabundant dark matter regions for heavier dark matter masses and suppressed effective interaction strengths.
}
\label{fig:omega_vs_mass}
\end{figure}

The benchmark analysis confirms the expected interplay between interaction strength, annihilation efficiency, and relic abundance. In particular, benchmark configurations with suppressed effective portal couplings exhibit inefficient annihilation and consequently large relic abundance, while simultaneously remaining consistent with present direct-detection bounds. Conversely, stronger effective interaction strengths enhance annihilation into electroweak final states such as $W^+W^-$, $ZZ$, and $hh$, thereby reducing the relic abundance but increasing the spin-independent dark matter--nucleon scattering cross section.

These results qualitatively reproduce the same tension structure observed throughout the numerical parameter scans, namely the competition between interaction suppression required by flavor constraints and efficient annihilation required for viable dark matter relic density. The benchmark validation therefore provides an independent numerical confirmation of the geometric compression of the viable parameter space discussed in the previous sections.

\begin{table}[t]
\centering
\begin{tabular}{c c c c c}
\hline
Benchmark & $M_{\rm DM}$ (GeV) & $\lambda_{SH}$ & $\Omega h^2$ & DD Status \\
\hline
BP1 & 1200 & 0.035 & 2.79 & Allowed \\
BP2 & 1000 & 0.08 & 0.407 & Excluded \\
\hline
\end{tabular}
\caption{
Representative benchmark configurations analyzed using \texttt{micrOMEGAs~6.3.0}. The benchmarks illustrate the interplay between annihilation efficiency, relic abundance, and direct-detection constraints in the effective dark matter framework.
}
\label{tab:dmbenchmarks}
\end{table}

\subsection{Texture Suppression and DM Viability}

An important question concerns whether the approximate texture suppressions identified earlier can alleviate the dark matter--LFV tension.

\begin{figure}[H]
\centering
\includegraphics[width=0.8\textwidth]{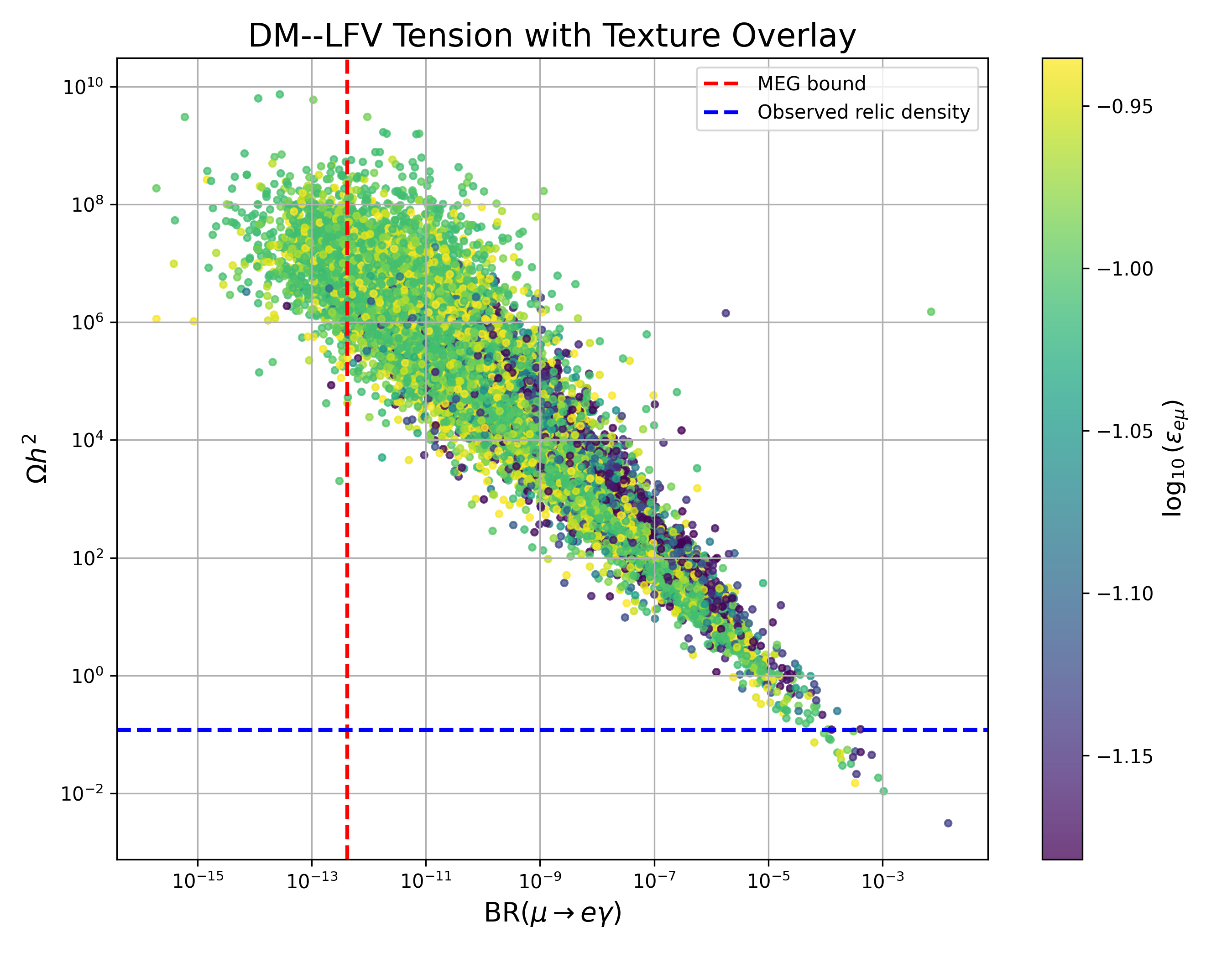}
\caption{
Dark matter relic density versus $\mathrm{BR}(\mu\to e\gamma)$ colored by the suppression parameter $\epsilon_{e\mu}$. Approximate flavor suppressions mildly soften the LFV tension but do not generically restore simultaneous dark matter and LFV viability.
}
\label{fig:dm_texture_overlay}
\end{figure}

To investigate this issue, we overlay the suppression parameter $\epsilon_{e\mu}$ onto the DM--LFV plane. The resulting distribution is shown in Fig.~\ref{fig:dm_texture_overlay}.

The figure reveals that smaller values of $\epsilon_{e\mu}$ tend to populate somewhat more LFV-favored regions. This confirms that emergent flavor suppressions partially reduce flavor-violating amplitudes through destructive interference.

However, the effect is comparatively modest. Approximate texture suppression alone does not generically restore simultaneous dark matter and LFV viability. Instead, the overall Yukawa magnitude continues to dominate the annihilation efficiency required for relic-density consistency.

This observation is conceptually important because it demonstrates that emergent textures soften but do not eliminate the underlying geometric tension.

\subsection{Phase Structure of Viable Parameter Space}

To further investigate the topology of the tension, we study the parameter-space distribution in the
\begin{equation*}
(M_1,\Delta m)
\end{equation*}
plane.

\begin{figure}[H]
\centering
\includegraphics[width=0.85\textwidth]{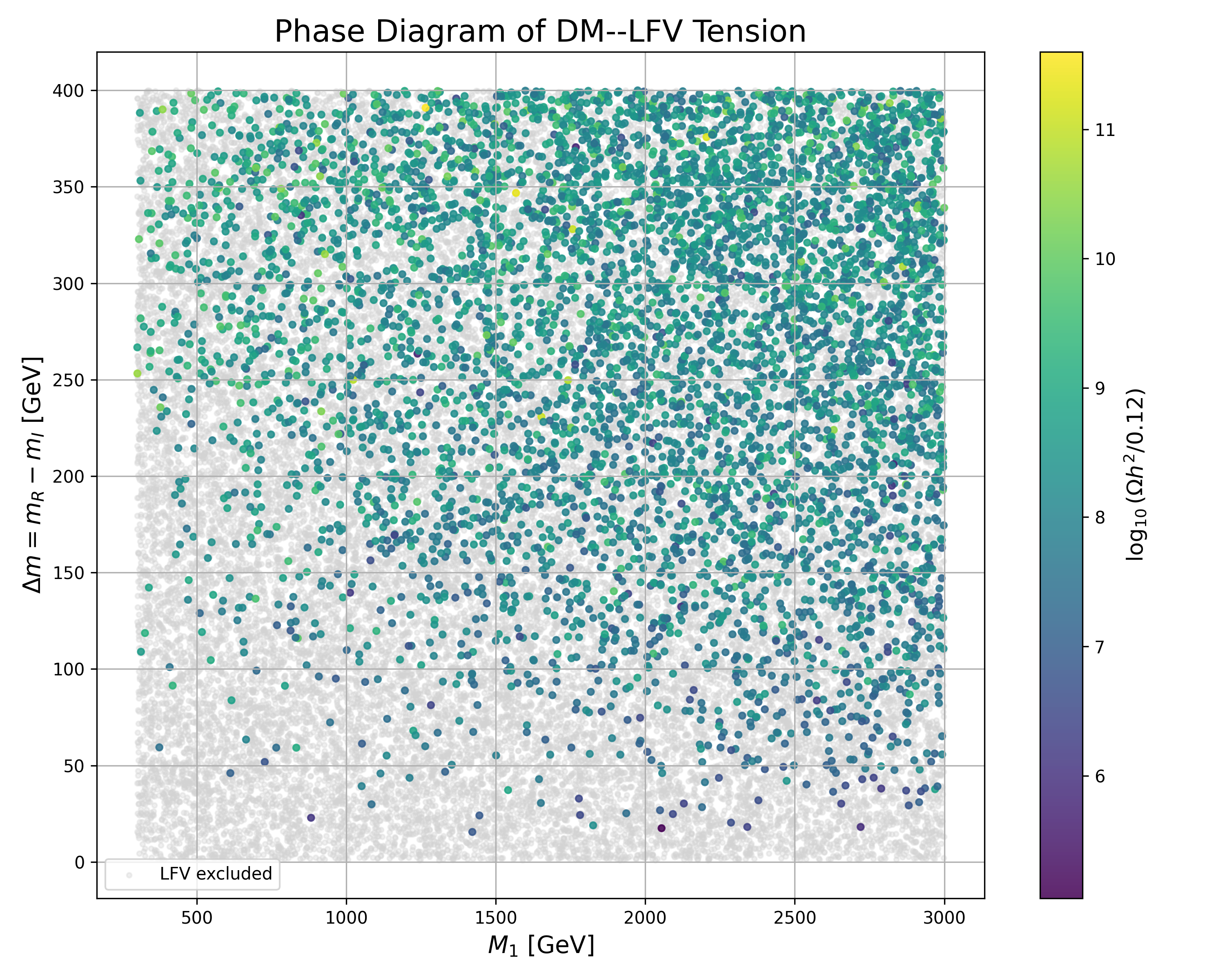}
\caption{
Phase diagram in the $(M_1,\Delta m)$ plane. Colored points satisfy LFV constraints and are shaded according to the relic-density ratio $\Omega h^2/0.12$. The figure demonstrates that LFV-safe regions generically populate strongly overabundant dark matter regimes across wide regions of parameter space.
}
\label{fig:phase_diagram}
\end{figure}

Figure~\ref{fig:phase_diagram} shows the resulting phase diagram, where points are colored according to the normalized relic-density ratio
\begin{equation}
\log_{10}\left(
\frac{\ohm}{0.12}
\right).
\end{equation}

A striking feature of the phase diagram is that LFV-safe points appear throughout a broad region of parameter space, indicating that the obstruction is not caused by the absence of accessible flavor configurations.

Instead, the dominant problem is persistent dark matter overabundance:
\begin{equation*}
\ohm
\gg
0.12
\end{equation*}
across most LFV-safe regions.

This observation demonstrates that the minimal fermionic scotogenic framework is not merely locally constrained, but instead globally overconstrained by the simultaneous requirements of:
\begin{equation*}
\text{oscillations}
\oplus
\text{LFV}
\oplus
\text{dark matter}.
\end{equation*}

The phase diagram therefore reveals a genuine structural obstruction within the minimal model.

\subsection{Coannihilation Effects}

A natural possibility for alleviating the tension involves compressed spectra and coannihilation effects. To investigate this scenario, we performed dedicated scans enforcing near-degeneracy between the fermionic dark matter candidate and the inert scalar sector.

Although coannihilation partially enhances annihilation efficiency, our numerical analysis indicates that it does not generically restore simultaneous LFV and relic-density consistency within the minimal setup. The viable parameter space remains strongly constrained, and dark matter overabundance persists across most LFV-safe regions.

This result suggests that the observed tension is not merely an artifact of simplified benchmark choices, but rather a robust consequence of the underlying radiative flavor geometry.

\section{Emergent Hierarchy Invariant and Flavor Compression}

\subsection{Emergent Hierarchy Invariant}

To characterize the relative accessibility of texture suppressions across flavor space, we define the emergent hierarchy invariant
\begin{equation}
\mathcal{H}
\equiv
\frac{
\epsilon_{\mu\mu}
+
\epsilon_{\tau\tau}
}{
\epsilon_{e\mu}
+
\epsilon_{e\tau}
},
\end{equation}

where $\epsilon_{\alpha \beta}$ has benn defined in Eq.(\ref{eq:cancellation}). The quantity $\mathcal{H}$ measures the relative geometric obstruction between atmospheric diagonal sectors and off-diagonal electron sectors.

Large values,
\begin{equation*}
\mathcal{H} \gg 1,
\end{equation*} 

indicate that the atmospheric sectors remain unsuppressed while strong cancellations become dynamically accessible in the electron off-diagonal sectors. 
Conversely,
\begin{equation*}
\mathcal{H} \sim \mathcal{O}(1),
\end{equation*}
corresponds to comparatively uniform suppression accessibility across flavor space.

\begin{figure}[H]
\centering
\includegraphics[width=0.95\textwidth]{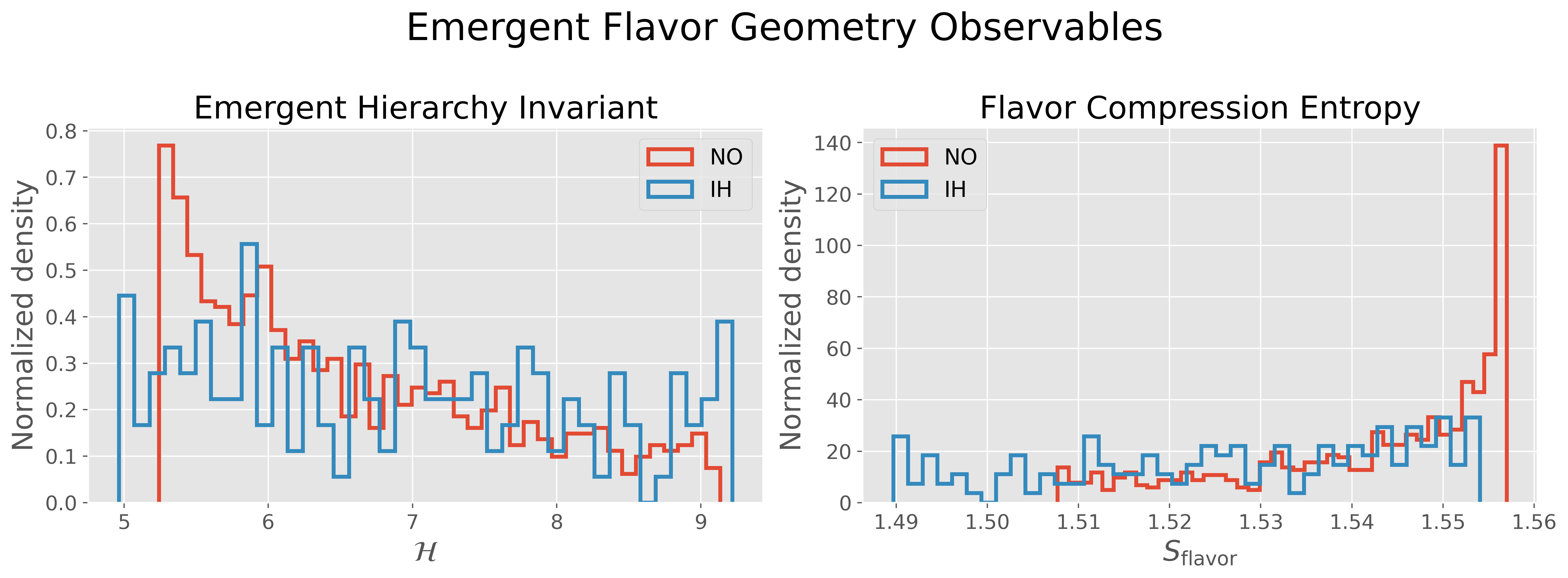}
\caption{
Normalized distributions of the emergent hierarchy invariant $\mathcal H$ and flavor-compression entropy $S_{\rm flavor}$ for viable normal-hierarchy (NH) and inverted-hierarchy (IH) solutions satisfying LFV constraints. The narrow entropy distributions indicate that viable radiative flavor structures populate a highly compressed submanifold of the full Casas--Ibarra parameter space. The broader hierarchy-invariant structure in IH further signals stronger geometric obstruction in realizing approximate texture suppressions.
}
\label{fig:hierarchy_entropy}
\end{figure}
The numerical scans reveal that LFV-safe regions of parameter space preferentially populate
\begin{equation}
\mathcal{H} \gg 1,
\end{equation}
particularly for normal neutrino hierarchy. 
This demonstrates that viable scotogenic parameter space dynamically favors hierarchical flavor compression rather than democratic texture accessibility.

\begin{figure}[H]
\centering
\includegraphics[width=0.95\textwidth]{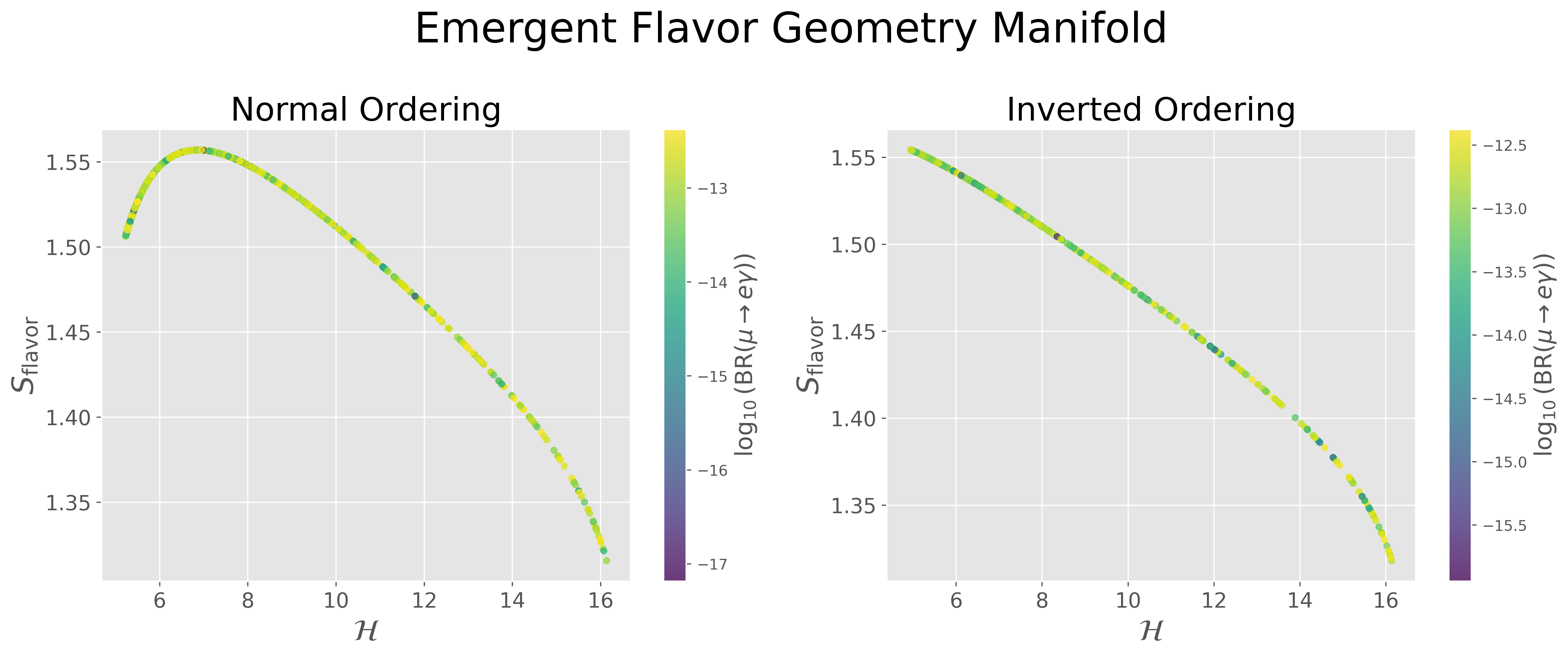}
\caption{
Emergent flavor-geometry manifold in the space of the hierarchy invariant $\mathcal H$ and flavor-compression entropy $S_{\rm flavor}$ for viable LFV-safe solutions. Both normal hierarchy (NH) and inverted hierarchy (IH) collapse onto highly localized low-dimensional trajectories rather than populating flavor space randomly. The stronger curvature and compression observed in IH indicate enhanced geometric obstruction in realizing approximate texture suppressions. Color coding denotes $\log_{10}({\rm BR}(\mu \to e\gamma))$.
}
\label{fig:flavor_geometry_manifold}
\end{figure}

\subsection{Flavor Compression Measure}

To further quantify the global distribution of suppression patterns across flavor space, we introduce the normalized flavor-compression weights

\begin{equation}
p_{\alpha\beta}
=
\frac{
\epsilon_{\alpha\beta}
}{
\sum_{\gamma\delta}
\epsilon_{\gamma\delta}
},
\end{equation}
which satisfy
\begin{equation*}
\sum_{\alpha\beta} p_{\alpha\beta} = 1.
\end{equation*}

Using these weights, we define the flavor-compression entropy
\begin{equation}
S_{\rm flavor}
=
-
\sum_{\alpha\beta}
p_{\alpha\beta}
\log p_{\alpha\beta}.
\end{equation}

The quantity $S_{\rm flavor}$ measures the effective geometric spread of viable texture structures across flavor space. Large values of $S_{\rm flavor}$ correspond to comparatively democratic texture accessibility, whereas smaller values indicate that viable parameter space becomes compressed into a restricted subset of flavor directions.

Our numerical analysis indicates that:
\begin{itemize}
\item normal hierarchy typically produces larger values of $S_{\rm flavor}$,
\item inverted hierarchy yields stronger geometric compression,
\item and LFV constraints systematically reduce the accessible flavor entropy.
\end{itemize}

These results demonstrate that phenomenological consistency conditions dynamically compress the viable flavor manifold into highly restricted geometric regions.

\section{Conclusions}

In this work, we investigated the emergence of approximate neutrino texture structures within the minimal fermionic scotogenic model using large-scale Casas--Ibarra parameter scans subject to oscillation consistency, lepton flavor violation constraints, and dark matter relic density requirements.

Unlike conventional texture-zero analyses, where flavor structures are externally imposed through symmetry assumptions, our approach treated texture suppressions as dynamically emergent properties of viable parameter space itself. To quantify these structures, we introduced normalized suppression measures for the radiatively generated neutrino mass matrix and analyzed their statistical distributions across the full Casas--Ibarra geometry.

Our analysis revealed several important features.

\begin{table}[t]
\centering
\begin{tabular}{c c c c c c}
\hline
Benchmark &
$M_{1}$ [GeV] &
$\Delta m$ [GeV] &
$\Omega h^2$ &
$\mathrm{BR}(\mu \to e \gamma)$ &
Status \\
\hline

BP1 & 1200 & 23.0  & 2.79 & $5.54 \times 10^{-14}$ & LFV-safe, DM overabundant \\

BP2 & 1681 & 267.5 & $2.43 \times 10^{4}$ & $3.08 \times 10^{-13}$ & Near MEG sensitivity \\

BP3 & 1852 & 87.8  & $4.43 \times 10^{4}$ & $7.28 \times 10^{-14}$ & LFV-safe, heavy DM \\

BP4 & 1277 & 205.8 & $4.50 \times 10^{4}$ & $7.82 \times 10^{-14}$ & Strong LFV suppression \\

BP5 & 1669 & 194.4 & $1.17 \times 10^{4}$ & $1.74 \times 10^{-12}$ & MEG-constrained \\

\hline
\end{tabular}
\caption{
Representative benchmark points from the combined dark matter and lepton-flavor-violation parameter-space scan. The table illustrates the interplay between relic-density generation, radiative flavor violation, and mass splitting effects in the viable emergent texture geometry. Benchmark points satisfying the current MEG bound correspond to
$\mathrm{BR}(\mu\to e\gamma) < 4.2 \times 10^{-13}$.
}
\label{tab:benchmark_dm_lfv}
\end{table}

First, approximate suppressions emerge nonuniformly across flavor space. In particular, the off-diagonal electron sectors
\begin{equation*}
M_{e\mu},
\qquad
M_{e\tau}
\end{equation*}
naturally develop significant suppressions within LFV-safe regions, while atmospheric diagonal sectors such as
\begin{equation*}
M_{\mu\mu},
\qquad
M_{\tau\tau}
\end{equation*}
strongly resist cancellation. These results demonstrate that viable scotogenic parameter space possesses a highly nontrivial emergent flavor geometry.

\begin{table}[t]
\centering
\begin{tabular}{c c c c c}
\hline
Texture Pattern & Hierarchy & LFV Status & DM Viability & Emergent Behavior \\
\hline
Approx.\ $M_{e\mu}\simeq0$ & NH & Allowed & Partial & Stable under scans \\
Approx.\ $M_{e\tau}\simeq0$ & NH & Constrained & Weak & Mild emergence \\
Approx.\ $M_{\mu\mu}\simeq0$ & IH & Disfavored & Poor & Strong LFV tension \\
Approx.\ $M_{\mu\tau}\simeq0$ & NH & Allowed & Good & Broad viable region \\
\hline
\end{tabular}
\caption{
Summary of representative emergent approximate texture structures identified in the numerical analysis. The table illustrates the interplay between flavor constraints, dark matter phenomenology, and viable geometric regions of parameter space.
}
\label{tab:emergent_textures}
\end{table}
Second, we showed that the accessible texture structures depend strongly on the neutrino mass ordering. Normal ordering admits substantially richer cancellation patterns and stronger approximate suppressions than inverted ordering. In particular, the $M_{ee}$ sector remains strongly protected against suppression within the inverted hierarchy due to the underlying interference structure of the neutrino mass eigenstates.

Third, we investigated the interplay between dark matter relic density and lepton flavor violation. Since both observables depend on the same Yukawa structures, viable parameter space becomes strongly constrained by a geometric tension:
\begin{equation*}
\text{LFV suppression}
\quad
\Longleftrightarrow
\quad
\text{dark matter overabundance}.
\end{equation*}

An additional important result concerns the robustness of the emergent flavor geometry. We demonstrated that the dominant texture structures remain qualitatively stable when comparing reduced and full Casas--Ibarra parametrizations. This indicates that the observed flavor hierarchies are genuine geometric properties of the radiative reconstruction framework rather than artifacts of specific parametrization choices.

Taken together, our results support a broader interpretation of flavor structure in radiative neutrino mass models:
\begin{center}
\emph{
approximate neutrino textures may emerge dynamically from the geometry of phenomenologically viable parameter space rather than from externally imposed flavor symmetries.
}
\end{center}

Within the scotogenic framework, oscillation reconstruction, Casas--Ibarra geometry, dark matter phenomenology, and lepton flavor violation collectively shape the accessible flavor structures of the neutrino mass matrix. The resulting flavor geometry is highly constrained, strongly ordering dependent, and deeply interconnected with dark matter dynamics.

The framework developed here opens several possible future directions. These include:
\begin{itemize}

\item scalar dark matter realizations of the scotogenic model,

\item direct detection and collider constraints,

\item leptogenesis within emergent flavor geometries,

\item generalized radiative flavor frameworks,

\item and the study of exact versus approximate texture manifolds.

\end{itemize}

More broadly, the results presented in this work suggest that emergent geometric structures may provide a useful organizing principle for understanding flavor hierarchies in radiative neutrino mass generation frameworks beyond the scotogenic model itself.

\end{document}